\documentclass[journal]{IEEEtran}

\usepackage{setspace}
\usepackage{cite,graphicx,amsmath,amssymb}
\usepackage{subfigure}
\usepackage{fancyhdr}
\usepackage{mdwmath}
\usepackage{mdwtab}
\usepackage{balance}
\usepackage{xcolor}
\usepackage{bm}
\usepackage{amsthm}
\usepackage{threeparttable}
\usepackage{algorithm}
\usepackage{algorithmic}
\usepackage{multirow}
\usepackage{flafter}
\usepackage{makecell}
\usepackage{diagbox}

\providecommand{\url}[1]{#1}
\allowdisplaybreaks
\begin{document}
\title{Near-Field Communications:\\ What Will Be Different?}

\author{
Yuanwei Liu,
Jiaqi Xu,
Zhaolin Wang,
Xidong Mu, 
and Lajos Hanzo
\thanks{Yuanwei Liu, Jiaqi Xu, Zhaolin Wang, and Xidong Mu are with the School of Electronic Engineering and Computer Science, Queen Mary University of London, London E1 4NS, UK, (email: \{yuanwei.liu, jiaqi.xu, zhaolin.wang, xidong.mu\}@qmul.ac.uk).}
\thanks{Lajos Hanzo is with the School of Electronics and Computer Science, University of Southampton, Southampton, SO17 1BJ, U.K. (e-mail: lh@ecs.soton.ac.uk).}
\vspace{-0.5cm}
}

\maketitle
\begin{abstract}
The design dilemma of \emph{“What will be different between near-field communications (NFC) and far-field communications (FFC)?”} is addressed from four perspectives. 1) From the  \emph{channel modelling perspective}, the differences between near-field and far-field channel models are discussed. A novel Green's function-based channel model is proposed for continuous-aperture antennas, which is contrasted to conventional channel models tailored for spatially-discrete antennas. 2) From the \emph{performance analysis perspective}, analytical results for characterizing the degrees of freedom and the power scaling laws in the near-field region are provided for both spatially-discrete and continuous-aperture antennas. 3) From the \emph{beamforming perspective}, far-field beamforming is analogous to a \emph{“flashlight”} that enables \emph{beamsteering}, while near-field beamforming can be likened to a \emph{“spotlight”} that facilitates \emph{beamfocusing}. As a further advance, a couple of new beamforming structures are proposed for exploiting the new characteristics of NFC.
4) From the \emph{application perspective}, new designs are discussed in the context of promising next-generation technologies in NFC, where our preliminary numerical results demonstrate that distance-aware target sensing and enhanced physical layer security can be realized in NFC. Finally, several future research directions of NFC are discussed.
\end{abstract}

\section{Introduction}
The emergence of revolutionary applications, such as extended reality (XR), digital twins, Metaverse, and holographic video, impose stringent requirements in terms of data rate, latency, reliability, coverage, and energy efficiency for next-generation (NG) wireless networks \cite{9349624}. Hence, efficient wireless technologies have to be conceived. On the one hand, motivated by the success of massive multiple-input multiple-output (MIMO) technology, the employment of extremely-large (XL)-MIMO and large-scale reconfigurable intelligent surfaces (RISs) has been proposed for improving the spectral efficiency. On the other hand, extremely high bandwidths are required for supporting the above services, but these are only available in the millimeter-wave (mmWave) and TeraHertz (THz) frequency bands. In this context, the importance of near-field communications (NFC) is becoming more pronounced, which presents both opportunities and challenges.

\begin{figure*}[t!]
  \begin{center}
  \includegraphics[width=1\linewidth]{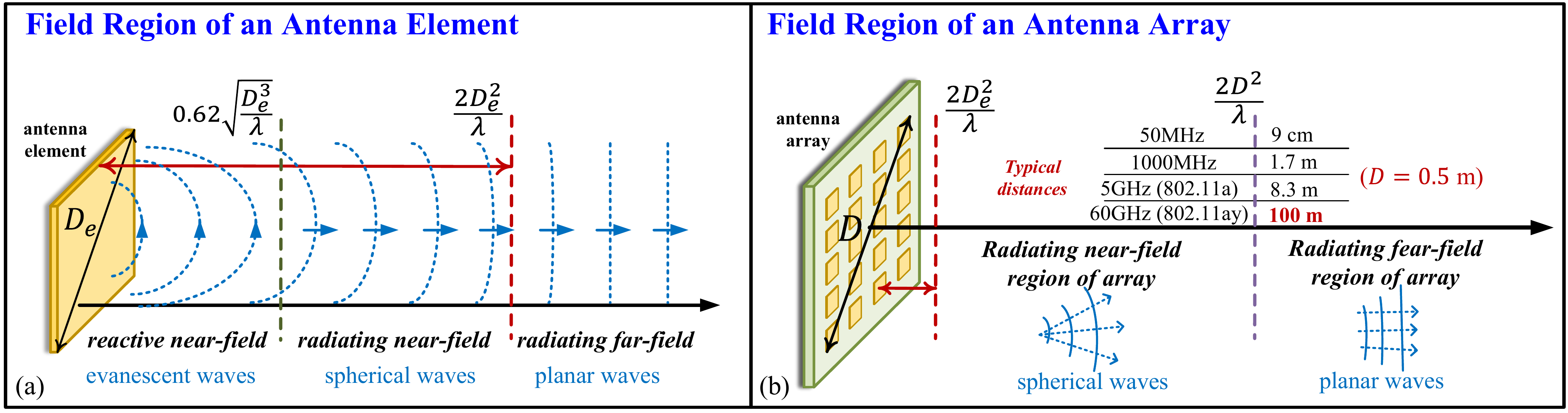}
      \caption{Illustration of field regions of (a) a continuous antenna element and (b) a spatially-discrete antenna array with respect to a point receiver.}
      \label{field_region}
  \end{center}
\end{figure*}

Specifically, according to electromagnetic (EM) and antenna theories, as shown in Fig. \ref{field_region}(a), the EM field radiated from the antennas can be divided into the following three regions having different properties:
\begin{itemize}
\item \textbf{\textit{Reactive near-field region}}, in which the electric and magnetic components of the field are not in-phase with each other. The energy of the EM field oscillates within this region, instead of being permanently removed from the transmitter. Therefore, in this region, non-propagating fields known as \emph{evanescent waves} dominate \cite {evan}. Because these waves decay rapidly with distance, they can only be leveraged within the reactive near-field region.
\item \textbf{\textit{Radiating near-field region}}, in which the electric and magnetic fields are perpendicularly in-phase with each other. Thus, the propagating waves emerge. However, angular field distribution is dependent upon the distance of receivers, resulting in the \emph{spherical wavefront}.
\item \textbf{\textit{Radiating far-field region}}, in which the signal paths between each point on the transmitter and the receiver can be regarded as parallel to each other. As a result, the angular field distribution is essentially independent of the distance between transmitters and receivers, resulting in the \emph{planar wavefront}. The decay rate of these far-field planar waves is the slowest among the aforementioned waves.
\end{itemize}
For a spatially-discrete antenna array, the boundaries between these regions are different for each antenna element and the entire antenna array. For a single antenna element, it can be regarded as a continuous radiating object. Therefore, the boundary between the reactive and radiating near-field regions is the start of the \textit{Fresnel region}, while the boundary between the near-field and far-field regions is known as the \textit{Rayleigh distance} \cite{balanis2015antenna}. The general expressions of these boundaries with respect to a point receiver\footnote{For a non-point receiver (e.g., multi-antenna receiver), each point on the receiver can be either located in the near-field or in the far-field region of the transmitter according to the boundary formulas evaluated in Fig. \ref{field_region}.} are given in Fig. \ref{field_region}(a), which are determined by the aperture of antenna elements and the signal wavelength. In practice, the aperture of antenna elements is typically on the order of sub-wavelength. As a result, the reactive and radiating near-field regions of antenna elements are negligible \cite{balanis2015antenna}. However, when it comes to XL-MIMO consisting of a large-scale spatially-discrete antenna array as shown in Fig. \ref{field_region}(b), the radiating near-field region of the entire array can be very large due to the joint effect of all antenna elements. It is worth noting that the reactive near-field region of a spatially-discrete antenna array is still negligible in practical communication scenarios, as the non-propagating evanescent waves only exist within a tiny region around each antenna element. Nonetheless, regarding XL-MIMO having continuous-aperture antennas, the reactive field region can also be relatively large due to the large continuous radiating surface \cite{evan}.

Naturally, near-field propagation always exists, but previous generations of wireless systems mainly relied upon far-field communications (FFC) since the corresponding near-field distance is quite limited. However, the rapid escalation in the number of antennas and frequencies in NG systems implies that we can no longer neglect NFC. Fig.~\ref{field_region}(b) lists some selected near-field distances at different frequencies for a spatially-discrete antenna array. Observed that even for a small antenna array, the Rayleigh distance at 60 GHz is on the order of hundreds of meters, which is comparable to the typical coverage of the 5G base station (BS). However, the investigation on NFC is still in its infancy, which raises the fundamental question, \textbf{\emph{“What will be different between NFC and FFC?”}}

To provide a complete picture of NFC, we answer the above question from four distinct perspectives, namely 1) channel modelling differences; 2) performance analysis differences; 3) beamforming differences; and 4) NG application differences; The rest of this article is organized as follows. 
In Section II, the differences between near-field and far-field channel models are discussed. In particular, a Green's function-based channel model is proposed for continuous-aperture antennas. 
In Section III, the performance analysis of near-field channels is carried out in terms of their power scaling laws and achievable degrees of freedom (DoFs).
In Section IV, the beamforming properties of NFC are identified. Then, a couple of beamforming structures are proposed for exploiting the new characteristics of NFC. 
Finally, in Section V, the unique features of NFC are discussed in an NG context.

\section{Differences in Near-Field Channel Modelling}
In this section, we will highlight the differences between near-field and far-field channel modelling. In addition, we provide a Green's function-based near-field channel model for continuous-aperture antennas.
In contrast to the far-field channel, the angular distribution of the near-field channel gain is a function of the distance between the transmitter and the observation point. Furthermore, the popular Friis formula~\cite{goldsmith2005wireless} is not valid in the near-field region. This necessitates the re-design of new near-field channel models. 
We mainly consider the radiating near-field region and contrast it to the radiating far-field region. 
Unless stated otherwise, “radiating near-field” is referred to as “near-field” for brevity.

\subsection{Near-Field Channel Models}
\subsubsection{New Opportunities}
The fundamental differences between near-field and far-field channel modelling open up new opportunities for NFC.
By taking into account the distance-dependent spherical wavefronts, new channel characteristics can be revealed and exploited. Here, we highlight a couple of favourable near-field channel characteristics. 
\textit{Firstly, \textbf{near-field channels have an enhanced degree of freedom (DoF)}}, which can be exploited for attaining a high multiplexing gain. The corresponding analytical results will be discussed in Section III.
\textit{Secondly, \textbf{Near-field channels exhibit improved beamfocusing capability compared to far-field channels}.} By exploiting the radiation distribution of near-field propagation, we can carry out \textit{beamfocusing} within the near-field region to achieve a higher channel gain than far-field \textit{beamsteering}. The differences between \textit{beamfocusing} and \textit{beamsteering} will be detailed in Section.~IV.

\subsubsection{New Challenges}
Near-field channels exhibit antenna-specific distances, which are different for each pair of transmit and receive antennas. 
As a result, it is non-trivial to accommodate these distances in the near-field channel models.
Different channel modelling methods are expected to be adopted for different types of transmitters.
We can broadly classify the existing transmitters into two categories, namely, \textit{spatially-discrete antennas} and \textit{continuous-aperture antennas}. To be more specific, conventional antenna arrays are spatially-discrete, because they typically have a spacing close to half-wavelength of their operating frequency \cite{balanis2015antenna}. By contrast, continuous-aperture antennas have antenna spacing that is significantly smaller than half-wavelength. This type of spatially continuous transmitters exploit densely populated antennas or metasurfaces to achieve a quasi-continuous aperture distribution~\cite{metasurface}.
For the specific structures of continuous-aperture antennas, it is challenging to obtain practical near-field channel models. To overcome this challenge, in the following, we proposed a Green's function-based near-field channel model for continuous-aperture antennas.

\subsection{Existing and Proposed Near-Field Channel Models}

\begin{table*}[!t]
\caption{Comparing fundamental performance limits of the radiating near-field and far-field regimes}
\label{tab:my-table}
\small
\begin{center}
\begin{tabular}{|c|c|c|c|}
\hline
\textbf{Field region}                                   & \textbf{Types of Antenna}  & \textbf{Power scaling law} & \textbf{Degrees of freedom} \\ \hline
\multirow{2}{*}{Radiating Far-Field}  & Spatially-discrete &  $\propto N_T$   &   1, free-space\\ \cline{2-4} 
                    & Continuous-aperture &  $\propto V_T$   &  1, free-space \\ \hline
\multirow{2}{*}{Radiating Near-Field}  & Spatially-discrete &  $\propto \sum_i^{N_T} d^{-2}_i$  &  $\min\{ N_T, N_R, \frac{2L_T^2L_R^2}{(d\lambda)^2}\}$  \\ \cline{2-4} 
                                 & Continuous-aperture &  $\propto \sum_i^N (\Delta V/d^2_i)$,~\cite{xu2022modeling}  & $\frac{2V_TV_R}{(d\lambda )^2 \Delta z_T \Delta z_R}$,~\cite{xu2022modeling} \\ \hline
\end{tabular}
\end{center}
In this table, $N_{T/R}$ is the number of transmitting/receiving antenna, $V_{T/R}$ is the volume of the transmitter/receiver, $d_i$ is the distance between the $i$th transmitting antenna and the receiver, $\Delta V$ denotes the maximum volume of the transmitter for the receiver to be located in its far-field, $N = V_T/\Delta V$, $\Delta z_T$ is the width of the transmitter, and $\Delta z_R$ is the width of the receiver~\cite{xu2022modeling}.
\end{table*}

\subsubsection{Models for Spatially-Discrete Antennas}
For the case of antenna arrays or patch-array-based RISs, the resultant near-field channel can be modelled by \textit{the complex-valued sum of all the far-field channels between each separate transmit antenna (TA) and the receive antennas (RA)}. This type of channel model is referred to as the non-uniform spherical wave (NUSW) model~\cite {9763525}. By doing so, the near-field channel between a pair of antenna arrays can be characterized by an $(N_R \times N_T)$-element channel matrix, where $N_R$ and $N_T$ are the numbers of RAs and TAs. The DoF of the near-field channel is given by the rank of this channel matrix, thus we can achieve full rank even without the presence of scatters.
In terms of channel state information (CSI) acquisition, near-field channels require higher precision to fully exploit their enhanced DoFs compared to far-field channels. For spatially discrete antennas, the NUSW channel models strike an appropriate trade-off between accuracy and complexity for most cases. For extremely large-scale MIMO systems, subarray-based estimation methods may be used~\cite{8949454}.

\subsubsection{The Model Proposed for Continuous-Aperture Antennas}
For the case of continuous-aperture antennas, the near-field channel modelling relies on Green's function method~\cite{danufane2020path}. Specifically, we propose a Green's function-based near-field channel for metasurface-based transmitters and simultaneously transmitting and reflecting reconfigurable intelligent surfaces (STAR-RISs), where Green's function can be regarded as the \textit{spatial impulse response} function between a continuous aperture transmitter and a near-field receiver~\cite{xu2022modeling}. 
By exploiting this model, an accurate volume-to-volume based model emerges, which is different from the conventional point-to-point model. As a benefit, the enhanced DoF of the near-field channel can be fully exploited, as seen in Section III.
Continuous-aperture antennas also require more accurate CSI compared to discrete antennas. Because their channel modeling relies on Green's function and spatial integration, the CSI acquisition of continuous-aperture antennas is a challenging topic for future investigation.

\section{Differences in Near-Field Performance Analysis}
In this section, we will provide insights concerning the power scaling law and the achievable DoF for near-field performance analysis and compare it to the far-field one. 
\subsection{Different Power Scaling Laws}
Here, we investigate how the received power scales as the size of transmitters increases, while both the transmit power and locations of transmitters and receivers are fixed. 
\subsubsection{New Opportunities and Challenges}
Within the near-field region, receivers are able to receive enhanced signal power from the transmitter. Specifically, by employing near-field beamfocusing, the power received within the near-field region can be substantially higher than that of the far-field region.
Nevertheless, power scales up slower in the near-field region than in the far-field region. As shown in Table~\ref{tab:my-table}, the received power increases linearly upon increasing the number of TAs in the far-field region,
while in the near-field region, it saturates. This is because the distances between the receiver and the edge of the transmitter ($d_i$) steadily increase. Eventually, when $d_i \to \infty$, the contribution of adding another TA tends to zero.
The saturation value depends on the geometric setup of the transceivers and can be calculated according to the near-field power scaling laws given in Table~\ref{tab:my-table}.

\subsubsection{Power Scaling Law Proposed for Continuous-Aperture Antennas}
Let us conceive the power scaling law for these continuous-aperture antennas. More particularly, we define a characteristic volume ($\Delta V$), which is the maximum transmitting volume for which the receiver can be regarded as a far-field receiver. 
As shown in Table~\ref{tab:my-table}, the near-field power is proportional to $\Delta V$ and inversely proportional to the square of the corresponding distance. We demonstrate that the ceiling of the proposed power scaling law for continuous-aperture antennas is higher than that of the spatially-discrete antennas.

\subsection{Different Achievable Degrees of Freedom}
\subsubsection{New Opportunities and Challenges}
NFC achieves a higher DoF than FFC because the spherical waves within the near-field region create different phase-shift and power levels for different links. If the antennas are spaced far enough from each other, the achievable DoFs of discrete antennas for the near-field region are equal to the minimum between $N_R$ and $N_T$.
\textit{\textbf{This indicates that spatial multiplexing can be used even without a rich scattering environment, which is one of the major benefits of NFC.}}
However, it is challenging to fully exploit the enhanced DoF supported by the near-field channel. As observed in Table~\ref{tab:my-table}, for communication at extremely high frequencies, the achievable DoF is limited both by the physical channel and by the number of TAs and RAs.
To overcome this challenge, below, we propose to exploit the near-field DoF for continuous-aperture antennas, which is not upper-bounded by the number of antennas.

\begin{figure*}[t!]
    \centering
    \includegraphics[width=0.8\linewidth]{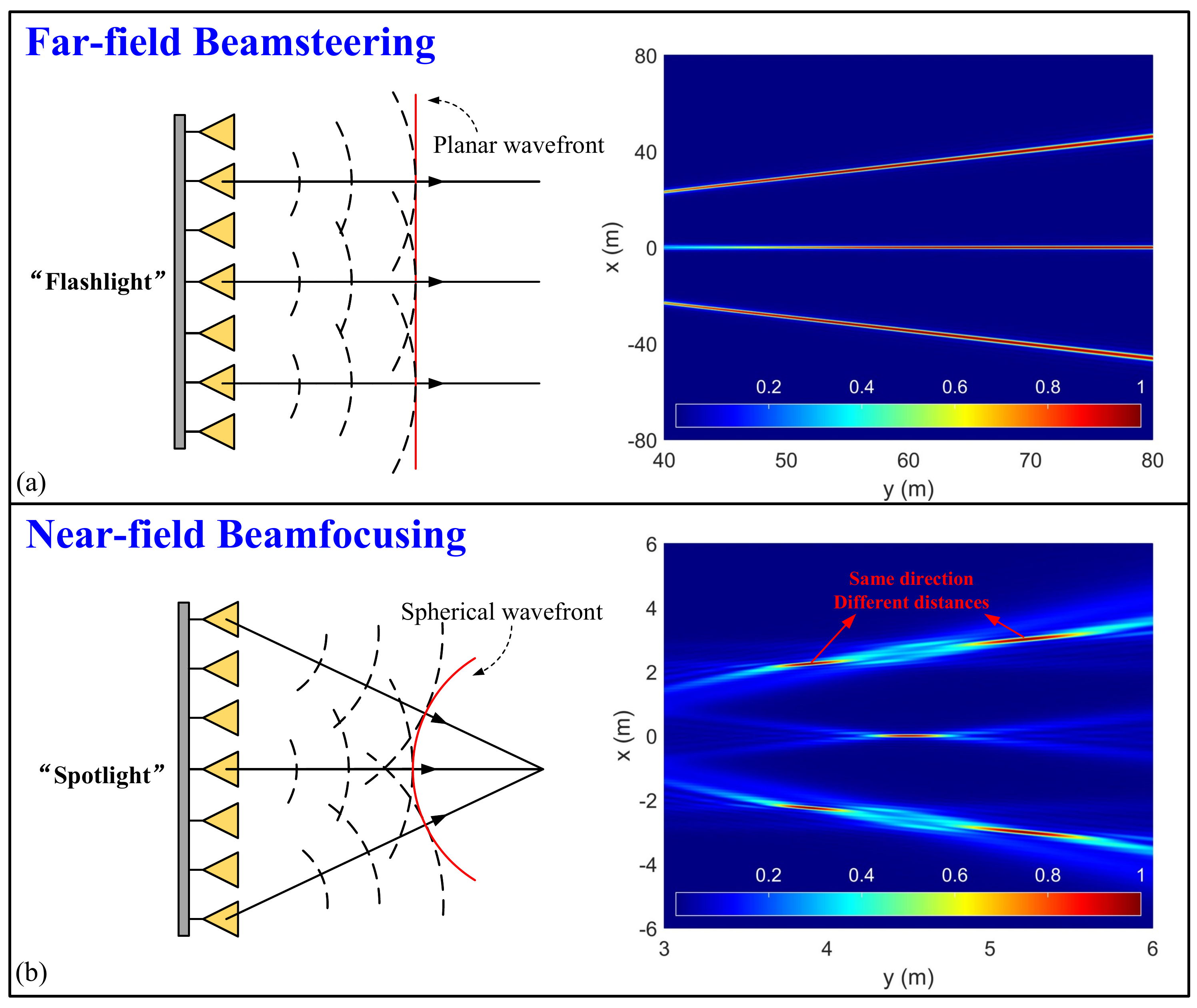}
    \caption{Beamforming in FFC and NFC. The beamsteering in FFC is like a \textbf{\emph{“flashlight”}} emitting light with a plane wavefront, while the beamfocusing in NFC is like a \textbf{\emph{“spotlight”}} emitting light with a spherical wavefront. The corresponding radiation patterns are illustrated on the right, where we consider a narrowband system with $128$ antennas operating at a frequency of $3$ GHz.}
    \label{beamforming}
\end{figure*}

\subsubsection{Achievable DoF for Continuous-Aperture Antennas}
As it may be inferred from Table~\ref{tab:my-table}, continuous-aperture antennas achieve a high DoF without using multiple antennas. This is because they can generate different current distributions within the TA aperture. At the same time, the receiver can collect received signal power at different locations of the RA aperture. Thus, multiple data streams can be transmitted between these antenna apertures by exploiting orthogonal electrical current distributions within these apertures.
The maximum number of parallel data streams that can be supported is equal to the number of communication modes~\cite{miller2000communicating}, i.e., DoFs of the near-field channel. 
According to the results derived in \cite{xu2022modeling}, the DoF of the near-field channel between two continuous apertures is proportional to the aperture size of the transceivers and inversely proportional to the square of the distance.
Consequently, by increasing the aperture sizes, the near-field channel can support high data rates and simultaneously achieve a high multiplexing gain.

\section{Differences in Beamforming}
In this section, we will provide a new look at beamforming in the near-field region. In particular, we first present the fundamental difference between near-field beamfocusing and far-field beamsteering, followed by identifying the new opportunities and challenges presented by near-field beamfocusing. Then, we propose a couple of new beamforming structures for NFC.

\subsection{Near-field Beamfocusing}

The difference in EM propagation characteristics results in different beamforming properties in these two regions, which can be illustrated by the following analogy. For FFC, the beamforming acts like a divergent \textbf{\emph{flashlight}}, as illustrated in Fig. 2(a), which is referred to as \textbf{\emph{beamsteering}}. By contrast, for NFC, the beamforming is like a more concentrated \textbf{\emph{spotlight}}, as illustrated in Fig. 2(b), which is referred to as \textbf{\emph{beamfocusing}}. To elaborate, the \emph{beamsteering} in FFC can only generate radiation patterns steered toward specific directions in the angular domain. By contrast, the \emph{beamfocusing} in NFC can be concentrated in locations determined by directions and distances in the polar domain \cite{9738442}. The additional distance dimension provided by the near-field beamfocusing opens new opportunities for advanced designs but has its own challenges.

\begin{figure*}[t!]
    \centering
    \includegraphics[width=0.8\linewidth]{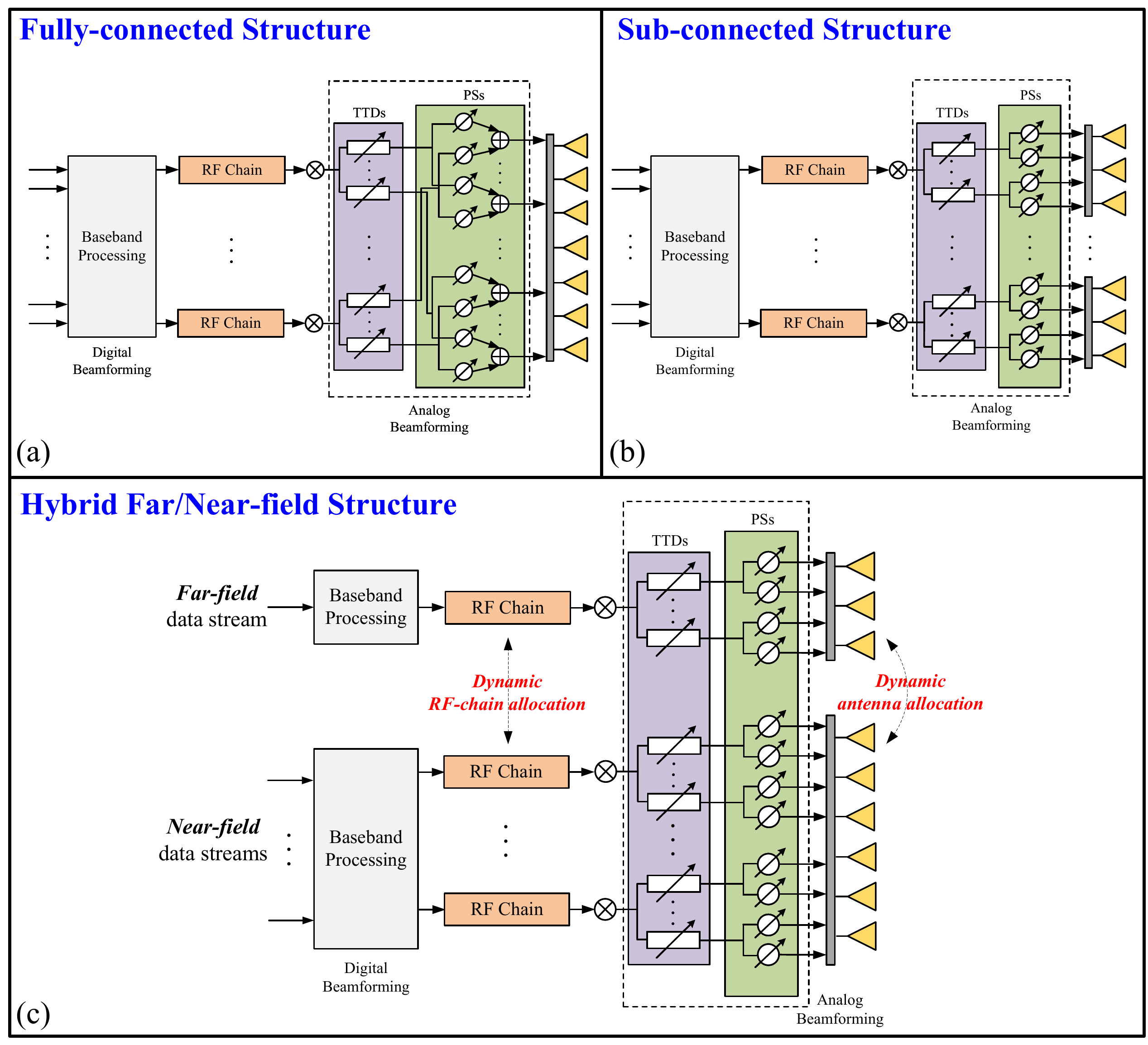}
    \caption{Beamforming structures for NFC: a) Fully-connected structure, b) Sub-connected structure, and c) Hybrid far/near-field structure.}
    \label{structure}
\end{figure*}

\subsubsection{New Opportunities}
\emph{Firstly}, similar to far-field beamsteering vectors, as the number of antennas tends to infinity, asymptotic orthogonality also holds for near-field beamfocusing vectors in the polar domain. Therefore, given the additional distance dimension, it is easier to create orthogonal not-interfering links for users in different locations via near-field beamfocusing than far-field beamsteering. \emph{Secondly}, near-field beamfocusing also makes wireless systems operating in high-frequency bands resistant to blockage. For far-field beamsteering, the receiver would be totally blocked in the presence of obstructions between the transmitter and the receiver. However, thanks to spherical wave propagation, it is highly likely that the majority of signals can still reach the intended receiver through near-field beamfocusing, as the different antenna elements at the transmitter can observe the receiver from different directions. \emph{Thirdly}, the beamforming strategy of the near-field region is not restricted to beamfocusing. Apart from the most popular spherical wavefront, some other irregular wavefronts can also be orchestrated in the near-field region, such as the Bessel beam wavefront \cite{singh2022wavefront}. The self-healing property of Bessel beams makes it possible to reconstruct directional beams after being blocked by obstacles. Therefore, compared to FFCs, NFC provides more opportunities for wavefront engineering and enables more flexible beamforming design.

\subsubsection{New Challenges}
However, the favourable properties of near-field beamfocusing also impose new challenges on the beamforming design. \emph{Firstly}, the near-field beamfocusing vector involves both angle and distance dimensions, which results in a much larger beamforming codebook than far-field beamsteering that involves only the angular dimension. Furthermore, in MIMO systems, the rank of near-field channels can be changed with the distance, which requires dynamic rank adaptation for near-field beamfocusing. These two issues dramatically increase the beamforming complexity in NFC. \emph{Secondly}, in high-frequency wideband systems, which is a common situation in NFC, the near-field channels can be strongly \emph{frequency-dependent} \cite{9398864}. These channels are not preferred by the \emph{frequency-independent} beamforming structure, such as the popular phase-shifter (PS) based analog beamformer, the beams generated by which in different frequencies may focus at different locations and cause \emph{beam split} effect \cite{9398864}. This effect causes the beam to be out of focus at a specific user position at different frequencies, thus resulting in significant performance loss. Compared to FFC, the beam split effect of NFC is more challenging to manage due to the non-linear phase in the near-field beamfocusing vector. \emph{Thirdly}, as the distance between the transmitter and receiver becomes large, exceeding the Rayleigh distance, the efficiency of near-field beamfocusing deteriorates and transitions to far-field beamsteering. In real-world scenarios, mobile users may exist in different fields or traverse between these fields, resulting in hybrid far-field and near-field issues. As a result, versatile adaptive designs are required for handling far-field beamsteering and near-field beamfocusing.

\subsection{Beamforming Structures Proposed for NFC}
Due to the grave \emph{frequency-dependence} of channels in wideband NFC, the conventional \emph{frequency-independent} PS-based analog beamforming is not suitable owing to its potential \emph{beam split} effect. As a remedy, it has been proposed to add an additional true-time-delayer (TTD)-based analog beamforming structure between the RF (radio-frequency) chains and the PS-based analog beamformer \cite{9398864}. In comparison to PSs, TTDs have the ability to delay a signal. According to the properties of the Fourier transform, these time delays manifest themselves as frequency-dependent phase shifts in the frequency domain. Therefore, the TTD-based structure facilitates frequency-dependent analog beamforming for mitigating the near-field beam split. The existing TTD-based hybrid beamformer requires that each RF chain is connected to all antennae via TTDs and PSs, which is referred to as a \textbf{\emph{fully-connected (FC) structure}} and illustrated in Fig. 3(a). Nevertheless, such a structure requires a large number of TTDs and PSs, which may still result in high power consumption, especially given that the power consumption of TTDs is much higher than that of PSs. Furthermore, this structure may also be unsuitable for hybrid far-field and near-field scenarios. To address these issues, we propose the \textbf{\emph{sub-connected (SC) structure}} and \textbf{\emph{hybrid far/near-field (HFN) structure}} in the following.

\subsubsection{Sub-connected Structure}

In contrast to the FC structure, each RF chain in the SC structure is only connected to a sub-array of the antenna array, as illustrated in Fig. 3(b). On the one hand, utilizing the SC structure is capable of reducing both the hardware complexity and power consumption by exploiting fewer hardware components than the FC structure. In particular, the number of PSs can be substantially reduced. Additionally, for smaller antenna sub-arrays, the beam split effect is less pronounced, hence requiring a reduced number of TTDs for each sub-array. On the other hand, the SC structure can also help reduce the beamforming complexity. This is because the communication links between each sub-array and the users can be approximated by FFC. Thus, low-complexity far-field channel models and state-of-the-art beamforming algorithms can be directly applied to each sub-array.

\subsubsection{Hybrid Far/Near-Field Structure}
Both SC and FC structures are subject to certain limitations. Firstly, whether a user is located in the near-field or far-field region depends upon the aperture size of the entire antenna array. Secondly, users located in the near-field region of the entire antenna array require complex near-field beamforming algorithms. However, this approach may not be suitable for cases where certain devices necessitate a low data rate, but have stringent requirements for latency. To overcome these limitations, we propose the HFN structure, as depicted in Fig. 3(c). In this structure, the entire antenna array can be dynamically partitioned into two separate sub-arrays of varying sizes. For the small sub-array, the original near-field users can be considered as far-field users. Due to the typically low rank of far-field channels, only a few or just a single RF chain is required and allocated for the small sub-array. Consequently, this sub-array can be used to serve delay-sensitive users relying on low-complexity far-field beamforming schemes. Simultaneously, the majority of the antennas and RF chains can be allocated to the larger sub-array to support users having high data rate requirements through NFC.

\section{Differences When Promising 6G Technologies Meet NFC} 
\begin{figure*}[t!]
    \begin{center}
        \includegraphics[width=0.8\linewidth]{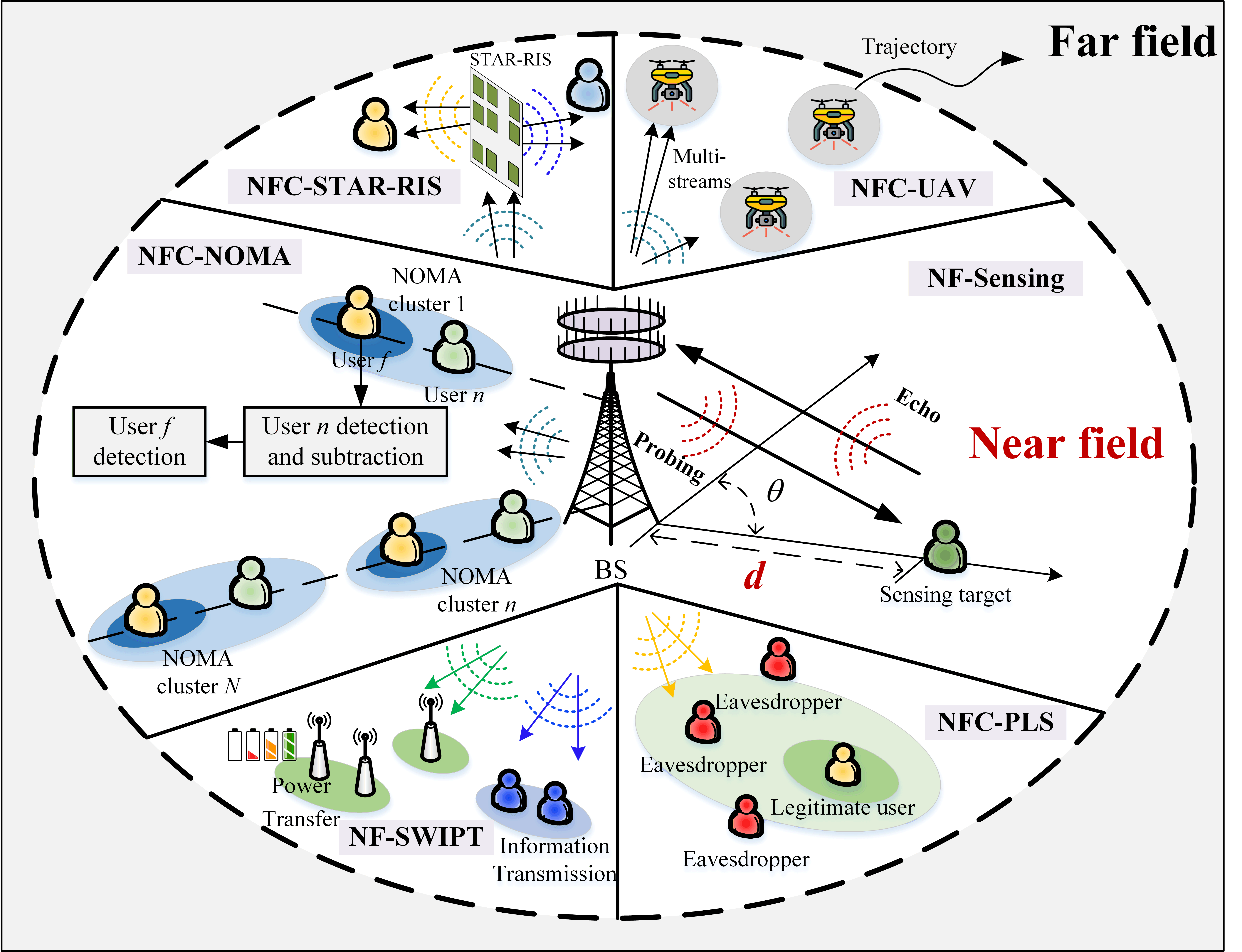}
        \caption{Illustration of promising 6G technologies in the near field.}
        \label{applications}
    \end{center}
\end{figure*}

Having introduced the fundamental beamforming structure, in this section, we will discuss the differences when employing promising 6G technologies in NFC, as illustrated in Fig. \ref{applications}.

\subsection{Near-Field NOMA Communications} 
The spherical wavefront of NFC paves the way for new non-orthogonal multiple access (NOMA) designs. By exploiting the new beamfocusing function of NFC, the users located far from the BS can have a higher effective channel gain than those located near the BS. These characteristics provide more flexibility for using NOMA in NFC. As shown at the left of Fig. \ref{applications}, a ``far-to-near'' successive interference cancellation (SIC) decoding order among users can be realized in near-field NOMA communications. Moreover, in contrast to conventional far-field NOMA, where users located in the same angular direction should be grouped into the same cluster, near-field beamfocusing allows the users in the same angular direction to be further grouped into different clusters. The advantages are that such designs reduce the total number of SIC operations employed at each user and simultaneously reduce the intra-cluster interference among users, thus realizing low-complexity NOMA communications in the near-field.

\subsection{Near-Field Sensing/ISAC} 
In near-field sensing, the spherical wave propagation relies on the polar coordinates of angle and distance. Thus, the sensing signal exhibits favourable structural characteristics, which can facilitate high-resolution holographic sensing. Fig. \ref{Sensing} characterizes the performance of near-field sensing through the MUltiple SIgnal Classification (MUSIC) algorithms \cite{wang2023near}, where joint angle and distance estimation is carried out. As it can be observed, with the aid of spherical wave propagation, the locations of the three targets can be accurately estimated, even though they are located in the same direction, which cannot be achieved in far-field sensing. However, the performance of near-field sensing degrades with the distance, since the spherical-wave propagation gradually transitions to plane-wave propagation. To solve this issue, hybrid far-field and near-field sensing is required. Alternatively, some advanced antenna architectures can be conceived for extending the near-field region, such as non-linear and holographic antenna arrays. Furthermore, since near-field can beneficially support both communication and sensing, it can also enhance the performance of integrated sensing and communication (ISAC) systems \cite{wang2023near}. In light of the new communication and sensing properties of the near field, it is necessary to redesign ISAC systems to enable mutually advantageous integration of its functions.

\subsection{Near-Field PLS} 
\begin{figure*}[!t]
    \centering
    \subfigure[Near-Field Sensing: Spectrum of MUSIC obtained by near-field sensing, where the BS is equipped with $512$ antennas and the three targets are located in the same direction with the locations of $(10 \text{ m}, 45^\circ)$, $(25 \text{ m}, 45^\circ)$, and $(40 \text{ m}, 45^\circ)$, respectively. Other simulation parameters can be found in \cite{wang2023near}]{\label{Sensing}
    \includegraphics[width=0.4\linewidth]{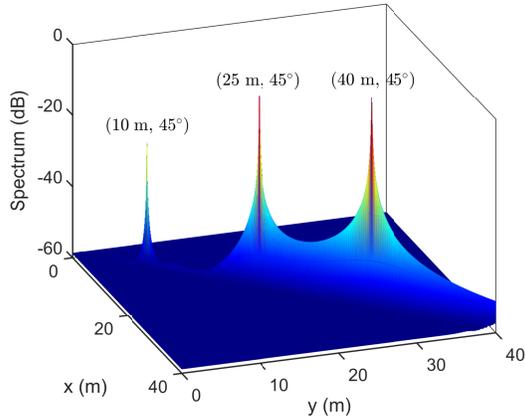}} \hspace{0.6cm}
    \subfigure[Neaf-Field PLS: Secrecy rate versus the distance between the eavesdropper and the BS, where the distance between the legitimate user and BS is fixed at 25 m, and the eavesdropper and the legitimate user have the same angular direction. Other simulation parameters can be found in \cite{Zheng_PLS}.]{\label{PLS}
    \includegraphics[width=0.4\linewidth]{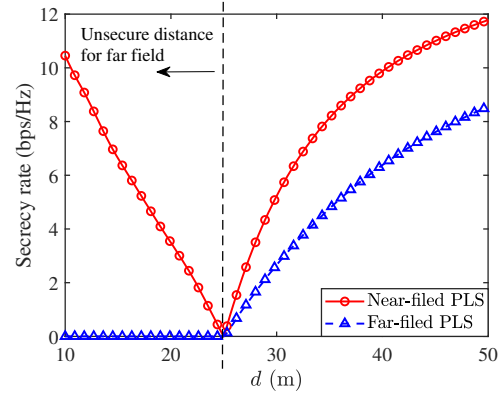}}
    \caption{Numerical results of NF-sensing and NFC-PLS.}\label{CS}
\end{figure*}

As shown at the bottom right of Fig. \ref{applications}, near-field beamfocusing can further reduce the information leakage in both the angle and location domains for further safeguarding the legitimate transmission in physical layer security (PLS). To demonstrate this, Fig. \ref{PLS} illustrates the secrecy rate versus the distance between the eavesdropper and the BS in both near-field and far-field communications \cite{Zheng_PLS}. It can be observed that NFC can further improve the secrecy rate even when the eavesdropper is closer to the BS than the legitimate user, which is fundamentally different from the PLS in FFC. It also shows that the PLS performance in NFC is mainly determined by the distance between the eavesdropper and the legitimate user rather than the distance between the eavesdropper and the BS in the FFC.

\subsection{Near-Field SWIPT} 
The near-field spherical wavefront provides new opportunities and challenges also for simultaneous wireless information and power transfer (SWIPT). As shown at the bottom left of Fig. \ref{applications}, on the one hand, near-field beamfocusing enables the BS to concentrate its the power on the specific location/region. Therefore, compared to far-field SWIPT, which dissipates energy in the angular domain, NF-SWIPT can substantially increase the efficiency of energy transfer. On the other hand, it is worth noting that despite the inter-functionality interference in NF-SWIPT can be further reduced by exploiting the near-field beamfocusing, it implies that dedicated beamfocusing vectors should be designed for power and information transmission. 

\subsection{STAR-RIS-aided NFC} 
In STAR-RIS-aided NFC, the near-field effects in the STAR-RIS are much more dominant and complicated than in the direct link. As shown at the top left of Fig. \ref{applications}, the propagation of both the incident, transmitted, and reflected signals can be within the near field, which imposes new challenges on both the STAR coefficient design and channel estimation. It is worth noting that STAR-RISs also provide enhanced flexibility for NFC. By employing the transmit/reflect mode-switching protocol, the number and the position of STAR elements working in the transmission-only or reflection-only mode can be adjusted, which results in different near-field boundaries in the transmission and reflection regions. This makes the near-field propagation surrounding the STAR-RIS eminently controllable.

\subsection{Near-Field Aerial Communications} 
In near-field aerial communications, the severe line-of-sight (LoS)-dominated interference can be efficiently mitigated by exploiting beamfocusing. As shown at the top right of Fig. \ref{applications}, a dedicated beamfocusing vector can be designed for serving each unmanned aerial vehicle (UAV) without causing significant inter-UAV interference. Moreover, in contrast to the LoS-dominated aerial channel in the far field, which is typical of low rank, higher ranks can be exploited in the near-field LoS-dominated aerial channel to support more data streams for each UAV. However, given the high mobility of UAVs, the trajectory design in mixed near/far-field communications is a challenging topic.

\section{Conclusions and the Road Ahead}
In this article, the fundamental differences between the emerging spherical wavefront-based NFC and the conventional plane wavefront-based FFC have been discussed in terms of channel modelling, performance analysis, beamforming structures, and integration with other NG technologies. 
However, Given that the investigation of NFC is still in its infancy, there are numerous open research problems calling for further research efforts. Some future research directions are exemplified below: 
\begin{itemize}
  \item \textbf{Stochastic Geometry Based Spatial Analysis for NFC}: Stochastic Geometry has been shown to be a powerful mathematical tool for far-field spatial analysis. In NFC, the achievable DoF is a distance-dependent parameter, which will usher in new research opportunities for stochastic geometry. New point processes and accurate 3-dimensional models have to be investigated to capture the spatial randomness of NFC. 
  \item \textbf{Channel Estimation and Beam Training for NFC}: Owing to the large antenna array and high-frequency bands used in NFC, the accurate estimation of the resultant high-dimensional CSI leads to potentially high pilot overheads. To this end, beam training is regarded as a promising solution to reduce the pilot overhead. Since near-field channels rely on both angular and distance information, a tailored polar-domain codebook and low-complexity near-field beam training protocols have to be developed, which requires future investigations.
  \item \textbf{Artificial Intelligence Aided NFC}: The complex signal propagation and extremely large numbers of antennas make the NFC design very challenging, where conventional optimization approaches suffer from potentially high complexity and low efficiency. Given their learning capability, diverse machine learning techniques can be employed in beamforming design, resource allocation, and channel estimation of NFC, which requires concerted community efforts.
\end{itemize}

\bibliographystyle{IEEEtran}
\bibliography{mybib}

\end{document}